\documentclass{article}
\usepackage[a4paper, portrait, margin=1.1811in]{geometry}
\usepackage[english]{babel}
\usepackage[utf8]{inputenc}
\usepackage[T1]{fontenc}
\usepackage{helvet}
\usepackage{etoolbox}
\usepackage{tikz}
\usepackage{pgfplots}
\usepgfplotslibrary{groupplots}
\usepackage{graphicx}
\usepackage{titlesec}
\usepackage{caption}
\usepackage{booktabs}
\usepackage{xcolor} 
\definecolor{table-green}{HTML}{0ec94d}
\usepackage[colorlinks, citecolor=table-green, linkcolor=red, urlcolor=blue]{hyperref}
\usepackage{caption}
\graphicspath{ {./images/} }
\usepackage{scrextend}
\usepackage{graphicx,subcaption}
\newcounter{lemma}

\newcounter{theorem}

\usepackage{tabularray}
\usepackage{amsmath} 
\usepackage{amsfonts} 
\UseTblrLibrary{booktabs}
\definecolor{gainsboro229}{RGB}{229,229,229}

\usepackage{fancyhdr}
\usepackage{lastpage}
\makeatletter
\patchcmd{\@maketitle}{\LARGE \@title}{\fontsize{16}{19.2}\selectfont\@title}{}{}
\makeatother

\usepackage{authblk}

\newsavebox\affbox
\author[1\dag]{\textbf{Kazi Shahriar Sanjid}}
\author[2\dag]{\textbf{Md. Tanzim Hossain}}
\author[3\dag]{\textbf{Md. Shakib Shahariar Junayed}}
\author[4*]{\textbf{Dr. Mohammad Monir Uddin}}
\affil[1,2,3]{Department of Electrical \& Computer Engineering, North South University, Dhaka-1229, Bangladesh}
\affil[4*]{Department of Mathematics and Physics, North South University, Dhaka-1229, Bangladesh}

\titlespacing\section{0pt}{11pt plus 4pt minus 2pt}{0pt plus 2pt minus 2pt}
\titlespacing\subsection{11pt}{11pt plus 4pt minus 2pt}{0pt plus 2pt minus 2pt}
\titlespacing\subsubsection{11pt}{11pt plus 4pt minus 2pt}{0pt plus 2pt minus 2pt}

\titleformat{\section}{\normalfont\fontsize{11}{15}\bfseries}{\thesection.}{1em}{}
\titleformat{\subsection}{\normalfont\fontsize{11}{15}\bfseries}{\thesubsection.}{1em}{}
\titleformat{\subsubsection}{\normalfont\fontsize{11}{15}\bfseries}{\thesubsubsection.}{1em}{}

\titleformat{\author}{\normalfont\fontsize{11}{15}\bfseries}{\thesection}{1em}{}

\title{\textbf{\huge Integrating Mamba Sequence Model and Hierarchical Upsampling Network for Accurate Semantic Segmentation of Multiple Sclerosis Legion}}
\date{}    

\begin{document}


\captionsetup[figure]{labelfont={bf},labelformat={default},labelsep=period,name={Figure }}	\captionsetup[table]{labelfont={bf},labelformat={default},labelsep=period,name={Table }}
\setlength{\parskip}{0.5em}
	
\maketitle
\thispagestyle{empty}	
\noindent\rule{15cm}{0.5pt}
	\begin{abstract}Integrating components from convolutional neural networks and state space models in medical image segmentation presents a compelling approach to enhance accuracy and efficiency. We introduce Mamba-HUNet, a novel architecture tailored for robust and efficient segmentation tasks. Leveraging strengths from Mamba-UNet and the lighter version of Hierarchical Upsampling Network (HUNet), Mamba-HUNet combines convolutional neural networks local feature extraction power with state space models long-range dependency modeling capabilities. We first converted HUNet into a lighter version, maintaining performance parity and then integrated this lighter HUNet into Mamba-HUNet, further enhancing its efficiency. The architecture partitions input grayscale images into patches, transforming them into 1-D sequences for processing efficiency akin to Vision Transformers and Mamba models. Through Visual State Space blocks and patch merging layers, hierarchical features are extracted while preserving spatial information. Experimental results on publicly available Magnetic Resonance Imaging scans, notably in Multiple Sclerosis lesion segmentation, demonstrate Mamba-HUNet’s effectiveness across diverse segmentation tasks. The model’s robustness and flexibility underscore its potential in handling complex anatomical structures. These findings establish Mamba-HUNet as a promising solution in advancing medical image segmentation, with implications for improving clinical decision-making processes. \\ \\
		\let\thefootnote\relax\footnotetext{
			\small $^{*}$\textbf{Corresponding author: {\color{table-green}monir.uddin@northsouth.edu.}}\\
                \small $^{\dag}$\textbf{These authors contributed equally to this work}}
		\textbf{\textit{Keywords}}: \textit{Medical image segmentation; State space model; Mamba; Multiple sclerosis; Convolutional neural networks}
	\end{abstract}
\noindent\rule{15cm}{0.4pt}

\section{Introduction}
Medical image segmentation, essential for diagnosing illnesses and planning treatments, is fundamental to various biomedical applications. Traditionally, manual segmentation has been labor-intensive, prompting a demand for automatic methods. Deep learning has revolutionized segmentation methodologies in recent years, significantly enhancing accuracy and efficiency across biomedical tasks \cite{bilic2023liver,heller2021state, hossain2023automated}.

Deep learning-based networks, like U-Net and DeepLab, excel in hierarchical feature extraction and are more parameter-efficient than fully connected networks \cite{lecun1995convolutional}, particularly the U-Net architecture, have exhibited superior performance in this domain \cite{long2015fully}. U-Net's symmetrical encoder-decoder style and skip connections facilitate effective feature extraction at different levels, enhancing information transformation efficiency \cite{ronneberger2015u}. Various enhancements to U-Net, incorporating advanced techniques like dense connections, residual blocks and attention mechanisms, have been explored, leading to modified versions for segmenting Computerized Tomography (CT), Magnetic Resonance Imaging (MRI) and ultrasound images \cite{ibtehaz2020multiresunet, li2018h, oktay2018attention, wang2021rar, zhang2020sau, zhou2019unet++}. 

Initially designed for natural language processing, transformers have been successfully adapted for image tasks, offering unique capabilities in capturing global information \cite{vaswaniattention}. Recent strides in computer vision owe much to the advancements inspired by natural language processing, notably exemplified by the Vision Transformer (ViT) and its introduction of a self-attention mechanism for image recognition, adept at modeling long-range dependencies \cite{dosovitskiy2020image}. Further adaptations like shift windows have expanded ViT's utility, leading to the emergence of the SwinTransformer, particularly suited for dense prediction tasks in computer vision \cite{liu2022swin, liu2021swin, liu2022video, xie2021self}. Within medical image segmentation, the fusion of ViT with U-Net architectures has birthed hybrid and pure ViT-based U-Nets, as evidenced by notable examples such as TransUNet and UNETR, harnessing ViT's feature learning prowess in various configurations \cite{chen2021transunet, hatamizadeh2022unetr}. Moreover, transformers, initially tailored for natural language processing, have found successful translation into image tasks, offering unique capabilities in capturing global information. This adaptability is strikingly evident in recent forays into 3D medical image segmentation, where UNETR incorporates ViT as its encoder. At the same time, SwinUNETR leverages the SwinTransformer, achieving promising outcomes through multi-scale feature extraction and fusion \cite{hatamizadeh2022unetr}. Nonetheless, the computational demands and reduced speed performance posed by the typically high resolution of 3D medical images remain noteworthy challenges for transformer-based methodologies.

Despite Transformers' proficiency in capturing long-range dependencies, their computational cost, especially for high-resolution biomedical images, remains a challenge \cite{ma2024u, xing2024segmamba}. State Space Models (SSMs), particularly Structured SSMs (S4), offer a promising solution with their efficiency in processing long sequences \cite{gu2023modeling, mehta2022long, wang2023selective}. Enhanced models like Mamba and VMamba further extend S4's applicability to computer vision tasks \cite{gu2023mamba, liu2024vmamba}. Due to the quadratic scaling of the self-attention mechanism with input size, the computational intensity of Transformers remains a challenge, particularly for high-resolution biomedical images. The SSMs, notably S4 and its enhanced versions like Mamba, have shown promise in efficiently modeling long sequences \cite{gu2021combining}. Mamba's selective mechanism and hardware optimization have surpassed Transformers on dense modalities such as language and genomics. SSMs offer a potential solution, exhibiting linear complexity concerning input size while establishing long-distance dependencies. Models like U-Mamba and SegMamba present hybrid SSM-CNN architectures for medical image segmentation tasks.

This paper proposes Mamba-HUNet, a novel biomedical image segmentation approach that integrates a hybrid dual-pathway multi-scale hierarchical upsampling network and a state space model. Mamba-HUNet is designed to effectively capture localized fine-grained features and long-range dependencies in biomedical images while ensuring linear scaling in feature size, thereby avoiding the quadratic complexity often associated with Transformers. Its self-configuring capability allows seamless adaptation to various datasets, enhancing scalability and flexibility across diverse biomedical segmentation tasks. Quantitative and qualitative evaluations on three distinct datasets demonstrate Mamba-HUNet's superior performance, surpassing Transformer-based networks significantly, thus paving the way for future advancements in network designs to model long-range dependencies in biomedical imaging efficiently. Additionally, we introduce a lightweight model suitable for deployment as a web application, an open API that accepts input images and returns disease labels, insights into the disease's etiology and suggested subsequent actions. This facilitates automatic annotation of new instances, benefiting the research community and represents a fundamental exploration into the potential of pure SSM-based models in medical image segmentation.

Key Contributions:
\begin{enumerate}
    \item Developed a lightweight version of the HUNet model \cite{sanjid4725376pixels}, maintaining performance parity, thereby enhancing computational efficiency in medical image segmentation tasks.
    \item Integrated the lightweight HUNet model into the Mamba architecture, resulting in superior performance compared to existing methods, as demonstrated through experimental evaluations.
    \item Achieved highly accurate predictions in multiple sclerosis lesion segmentation, a critical task in medical radiology, highlighting the practical significance and efficacy of the proposed Mamba-HUNet model in clinical applications. Additionally, we developed a user-friendly website with an API, facilitating seamless integration of the Mamba-HUNet model into radiologists workflow for improved diagnosis and treatment planning.
\end{enumerate}

\section{Preliminaries}
\subsection{State Space Models}
The SSMs are a class of systems that relate a one-dimensional function or sequence $u(t) \longmapsto y(t) \in \mathcal{R}$ and can be expressed as a linear time-invariant (LTI) continuous-time system of the form \cite{DBLP:journals/corr/abs-2111-00396}
\begin{equation}
\begin{aligned}
  \dot{x}(t) &= \textbf{A}x(t) + \textbf{B}u(t), \\
  y(t) &=  \textbf{C}x(t) +\textbf{D}u(t),
\end{aligned}
\label{sss}
\end{equation}

where $\textbf{A} \in \mathbb{R}^{n \times n}$ and $\textbf{B}\in\mathbb{R}^{n\times 1}$ and $\textbf{C} \in \mathbb{R}^{1\times n}$, are representing the system, input and output matrices respectively, of the system. The vectors  $x(t)$, $u(t)$ and $y(t)$, respectively known as state, input and output of the system.  These models possess advantageous features like linear computational complexity per time step and parallelized computation, which facilitate efficient training. Nonetheless, conventional SSMs tend to require more memory compared to equivalent convolutional neural networks (CNNs) and frequently face challenges such as vanishing gradients during the training process, limiting their broader utility in general sequence modeling. 

The S4 \cite{gu2021efficiently} enhances conventional SSMs by introducing structured patterns in the state matrix $A$ and employing a proficient algorithm. In particular, the state matrix is constructed and initialized using the High-Order Polynomial Projection Operator (HIPPO) \cite{gu2020hippo}, creating deep sequence models with enhanced capabilities for efficient long-range reasoning. S4, as an innovative network architecture, has outperformed Transformers \cite{vaswani2017attention} by a considerable margin in the demanding Long Range Arena Benchmark \cite{tay2020long}. Mamba introduces notable advancements in discrete data modeling using SSMs, particularly in domains such as text and genomics. Two primary enhancements distinguish Mamba from traditional SSMs approaches. Firstly, it incorporates an input-dependent selection mechanism, unlike conventional SSMs, invariant to time and input, enabling efficient information extraction tailored to specific inputs. This is achieved by parameterizing the SSMs parameters based on the input data. Secondly, Mamba introduces a hardware-aware algorithm that scales linearly with sequence length, facilitating recurrent model computation through scanning, thereby outperforming previous methods on modern hardware. The Mamba architecture, which integrates SSMs blocks with linear layers, is notably simpler and has exhibited state-of-the-art performance across various long-sequence domains such as language and genomics. This underscores its significant computational efficiency in both the training and inference stages.

\subsection{Discrete-time SSM: The Recurrent Representation}
When dealing with a discrete input sequence $(u_0, u_1, \ldots)$ instead of a continuous function $u(t)$, it's necessary to discretize equation \ref{sss} using a \textbf{step size $\Delta$}, which denotes the granularity of the input. Essentially, the inputs $u_k$ can be seen as samples taken from an underlying continuous signal $u(t)$, where $u_k = u(k\Delta)$. 

To discretize the continuous-time State Space Model (SSM), we adopt the bilinear method, as established in previous studies such as \cite{tustin1947method}. This method transforms the state matrix $\textbf{A}$ into an approximation $\mathbf{\overline{A}}$. The discrete SSM is

\begin{equation}
\begin{aligned}
  x_k &= \mathbf{\overline{A}}x_k + \mathbf{\overline{B}}u_k\\
  \mathbf{\overline{A}} &= (\mathbf{\overline{I}} - \Delta/2.\mathbf{\overline{A}})^{-1}(\mathbf{\overline{I}} + \Delta/2.\mathbf{\overline{A}})\\
  y_k &= \mathbf{\overline{C}}x_k\\
  \mathbf{\overline{B}} &= (\mathbf{\overline{I}} - \Delta/2.\mathbf{\overline{A}})^{-1}\Delta \mathbf{\overline{B}}\\
  \mathbf{\overline{C}} &= \mathbf{\overline{C}}
\end{aligned}
\label{fds}
\end{equation}

In Equation \ref{fds}, we now treat $u_k$ as a sequence-to-sequence mapping to $y_k$, departing from its previous function-to-function interpretation. Additionally, the state equation now recurs in $x_k$, enabling the discrete State-Space Model (SSM) to be computed like a Recurrent Neural Network (RNN). Specifically, $x_k \in \mathbb{R}^N$ serves as a hidden state with a transition matrix denoted by $\mathbf{\overline{A}}$.

Throughout this paper, we denote the discretized SSM matrices by $\mathbf{\overline{A}}, \mathbf{\overline{B}}, \ldots$, which are determined by Equation \ref{fds}. It's worth noting that these matrices depend not only on $\textbf{A}$ but also on a step size $\Delta$. For the sake of brevity, we omit explicit mention of this dependence in our notation when it's understood.

\subsection{Training SSMs: The Convolutional Representation}

The recurrent SSM referenced as \ref{fds} poses practical challenges for training on contemporary hardware because of its sequential nature. However, there exists a widely acknowledged link between linear time-invariant (LTI) SSMs like \ref{sss} and continuous convolutions. Consequently, \ref{fds} can be reformulated as a discrete convolution. 

To simplify, assume the initial state is $x_{-1} = 0$. Then unrolling \ref{fds} explicitly yields

\begin{align*}
  x_0 &= \mathbf{\overline{B}}u_0 & x_1 &= \mathbf{\overline{AB}}u_0 + \mathbf{\overline{B}}u_1 & x_2 &= \mathbf{\overline{A}}^2\mathbf{\overline{B}}u_0 +\mathbf{\overline{AB}}u_1 + \mathbf{\overline{B}}u_2 & \hdots\\
  y_0 &= \mathbf{\overline{CB}}u_0 & y_1 &= \mathbf{\overline{CAB}}u_0 + \mathbf{\overline{CB}}u_1 &  y_2 &= \mathbf{\overline{CA}}^2\mathbf{\overline{B}}u_0 + \mathbf{\overline{CAB}}u_1 + \mathbf{\overline{CB}}u_2 & \hdots
\end{align*}

This can be vectorized into a convolution \ref{jytj} with an explicit formula for the convolution kernel \ref{fghdf}.

\begin{equation}
\begin{aligned}
  y_k &= \mathbf{\overline{CA}}^k\mathbf{\overline{B}}u_0 + \mathbf{\overline{CA}}^{k-1}\mathbf{\overline{B}}u_1 + \hdots + \mathbf{\overline{CAB}}u_{k-1} + \mathbf{\overline{CB}}u_k,\\
  y &= \mathbf{\overline{K}} * u.
\end{aligned}
\label{jytj}
\end{equation}

\begin{equation}
\begin{aligned}
  \mathbf{\overline{K}}\in \mathbb{R}^L := \mathcal{K}_L(\mathbf{\overline{A}}, \mathbf{\overline{B}}, \mathbf{\overline{C}}) := (\mathbf{\overline{CA}}^i\mathbf{\overline{B}})_{i \in [L]} = (\mathbf{\overline{CB}}, \mathbf{\overline{CAB}} \hdots, \mathbf{\overline{CA}}^{L-1}\mathbf{\overline{B}}).
\end{aligned}
\label{fghdf}
\end{equation}
In other words, equation \ref{jytj} represents a solitary (non-circular) convolution and can be efficiently calculated using FFTs, as long as $\mathbf{\overline{K}}$ is provided. However, determining $\mathbf{\overline{K}}$ in \ref{fghdf} is complex and forms the core of our technical advancements in {\color{red}Section 3}. We refer to $\mathbf{\overline{K}}$ as the \textbf{SSM convolution kernel} or filter.

\begin{figure*}[t]
     \centering
     \includegraphics[width=\textwidth]{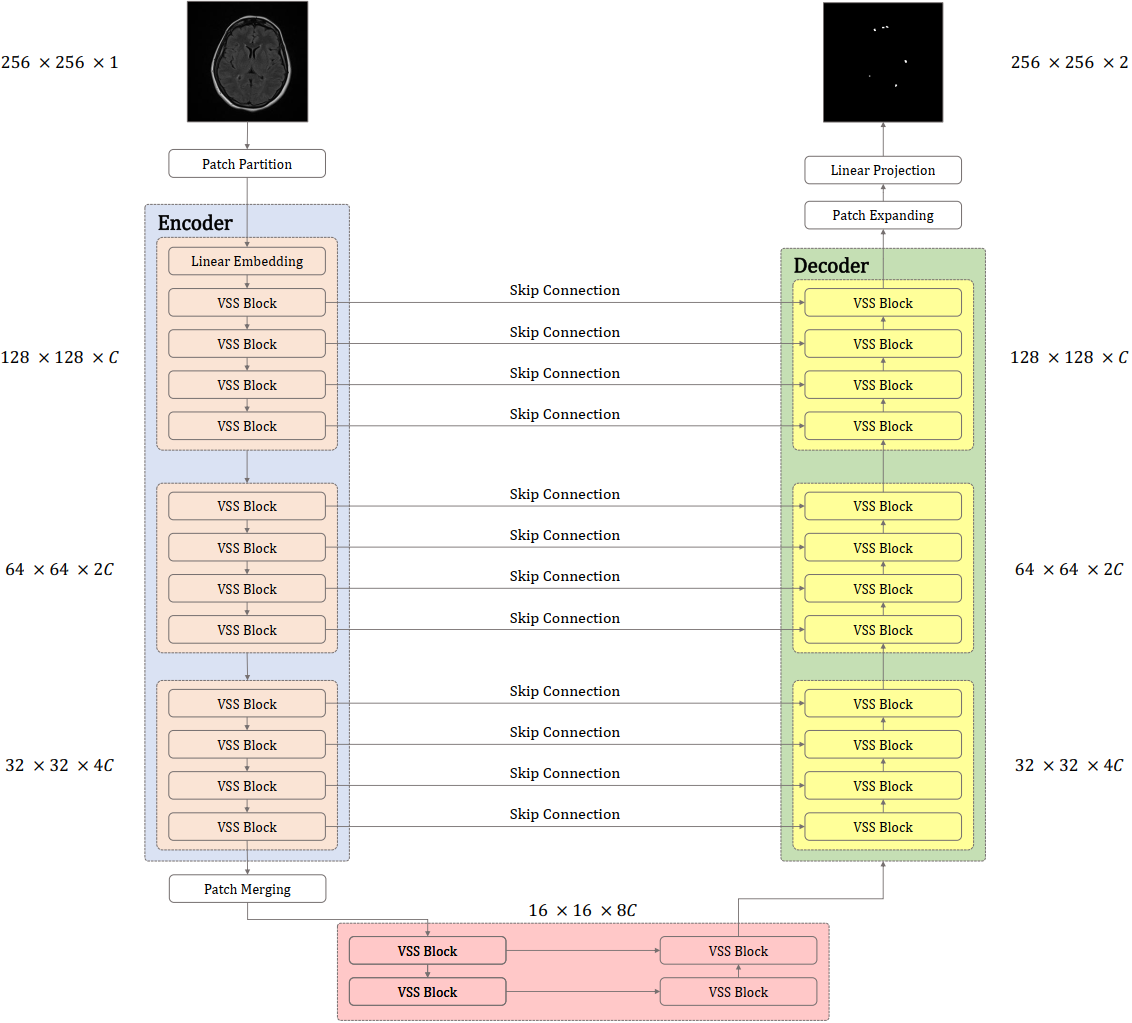}
     \caption{Overview of the Mamba-HUNet architecture.}
    \label{fig:Mamba-HUNet}
\end{figure*}

 \begin{figure*}[t]
     \centering
     \includegraphics[width=0.5\textwidth]{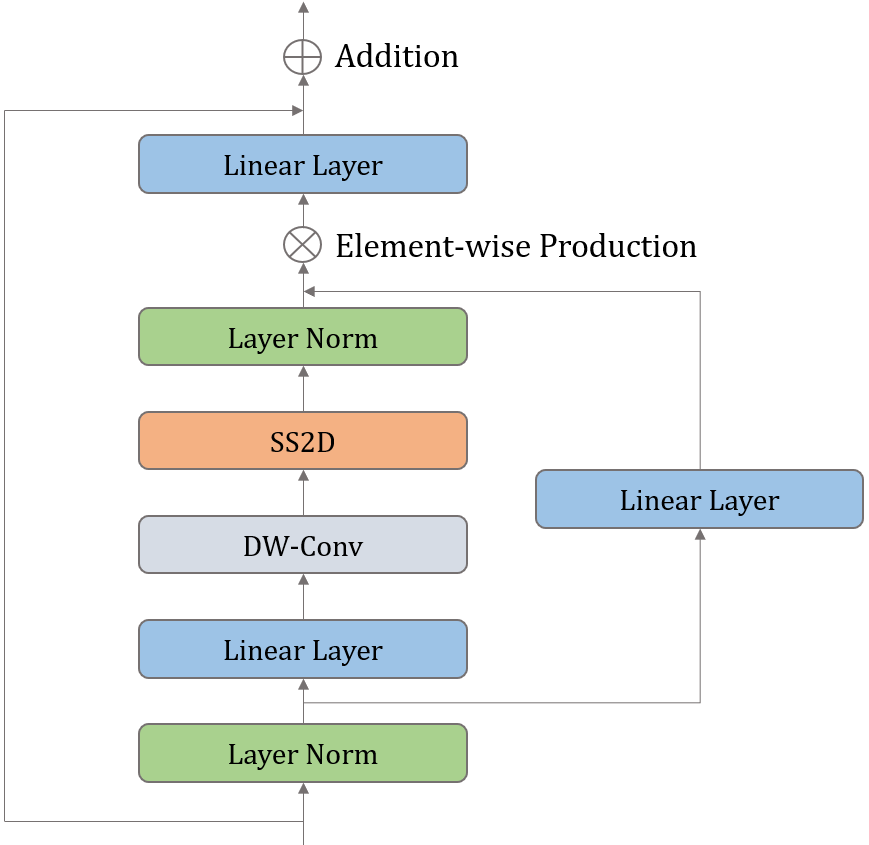}
     \caption{VSS block architecture}
    \label{fig:vss-block}
\end{figure*}

\section{Mamba-HUNet}
The selective space architecture (SSA) is a foundational component in the model, orchestrating a series of operations to prepare input tensors and parameters for selective scan processing. The SSA optimally structures the data to leverage selective attention mechanisms efficiently through chunking, convolution and reshaping. Discretizing the continuous-time selective SSM involves converting the model's representation from a continuous to a discrete-time domain. This process samples the continuous-time signal at specific intervals, resulting in a sequence of discrete-time states. This discretization allows the model to approximate its state matrix, enabling sequence-to-sequence mapping and recurrent computations like those performed by recurrent neural networks (RNNs). Essentially, it empowers the model to process input sequences over time and maintain internal states across different time steps, mirroring the behavior of RNNs. The selective scan operation within the 2D-selective-scan (SS2D) model further enhances this process, enabling selective aggregation of informative features while managing computational complexity effectively. By integrating bidirectional processing information and selectively attending to key spatial locations and feature channels, the selective scan operation enriches the model's ability to capture intricate patterns and dependencies within the data. This selective aggregation mechanism enhances contextual understanding and boosts computational efficiency by minimizing redundant computations.
The integration of selective scan with the HUNet layer in the subsequent stages of the model ensures seamless feature integration across different architectural components

\begin{figure*}[t]
     \centering
     \includegraphics[width=\textwidth]{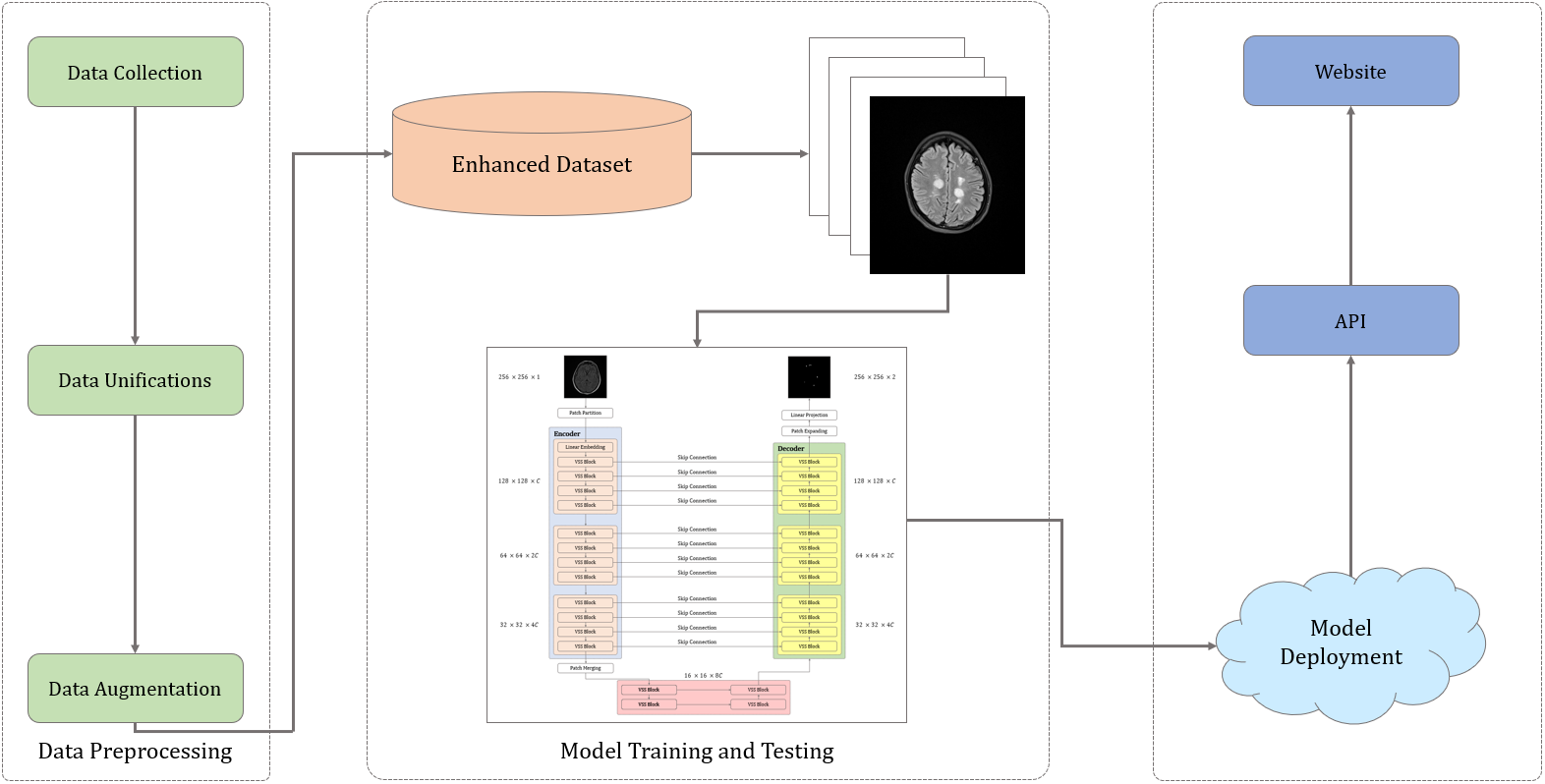}
     \caption{Overall workflow}
    \label{fig:workflowjut}
\end{figure*}

Incorporating a lighter version of the HUNet model into the methodology involves adapting the architecture to reduce computational complexity and enhance efficiency without compromising performance. This consists of modifying the original HUNet architecture by implementing strategies such as reducing the number of parameters, adjusting dropout rates and simplifying convolutional layers while maintaining accuracy in segmentation tasks. By integrating insights from the lighter HUNet version, the Mamba-HUNet architecture can effectively handle diverse datasets and complex anatomical structures while offering improved computational efficiency and streamlined processing. This adaptation seamlessly integrates into the website API, providing a robust solution for medical image segmentation tasks with enhanced efficiency and performance, as detailed in Figure \ref{fig:workflowjut}.

The proposed lighter Mamba-HUNet architecture integrates features from the Mamba-UNet \cite{wang2024mamba} and HUNet models to address the complexities of medical image segmentation tasks. The architecture begins by partitioning the input 2D grey-scale image into patches. These patches are then transformed into a 1-D sequence, facilitating efficient processing. An initial linear embedding layer adapts the feature dimensions, after which the data flows through multiple Visual State Space (VSS) blocks and patch merging layers. Inspired by Mamba-UNet, the VSS blocks leverage depth-wise convolution and softmax activation for feature extraction, avoiding positional embedding to maintain a streamlined structure. The encoder generates hierarchical features at different resolutions, preserving spatial information while reducing the token count and doubling feature dimensions through patch merging layers. Each encoder level utilizes skip connections to blend multi-scale features and maintain spatial detail. The decoder mirrors the encoder's structure, employing patch-expanding layers for upsampling and enhancing spatial resolution. The encoder and decoder incorporate VSS blocks to facilitate feature learning and reconstruction. The final segmentation masks are produced through convolutional layers, with the number of output channels corresponding to the classes to be segmented. Batch normalization and dropout regularization are applied throughout the architecture to enhance model stability and prevent overfitting. Incorporating insights from HUNet, the Mamba-HUNet architecture integrates additional convolutional layers and transposed convolutional layers within the upsampling block, bolstering the model's ability to capture intricate features and facilitate accurate segmentation. These enhancements empower the model to handle diverse datasets and complicated anatomical structures adeptly. By combining the strengths of both architectures, Mamba-HUNet offers a robust solution for medical image segmentation tasks, with the flexibility to handle diverse datasets and complex anatomical structures effectively.

\section{Experimental Procedures}
\subsection{Dataset Description}
\begin{figure}
  \begin{subfigure}{0.325\columnwidth}
  \includegraphics[width=\textwidth, height=\textwidth]{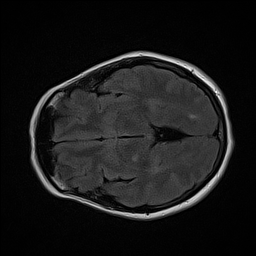}

  \end{subfigure}
  \hfill
  \begin{subfigure}{0.325\columnwidth}
  \includegraphics[width=\textwidth, height=\textwidth]{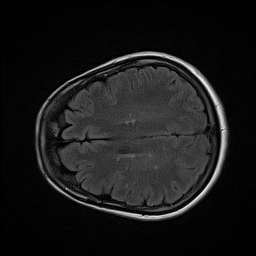}
  
  \end{subfigure} 
  \hfill
  \begin{subfigure}{0.325\columnwidth}
  \includegraphics[width=\textwidth, height=\textwidth]{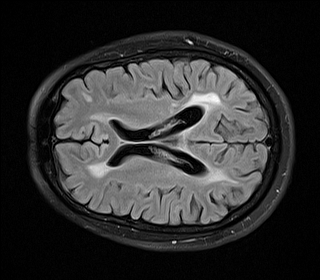}

  \end{subfigure} 
  \begin{subfigure}{0.325\columnwidth} 
  \includegraphics[width=\textwidth, height=\textwidth]{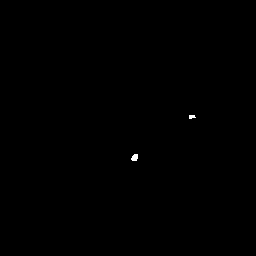} 

  \end{subfigure}  
  \hfill 
  \begin{subfigure}{0.325\columnwidth} 
  \includegraphics[width=\textwidth, height=\textwidth]{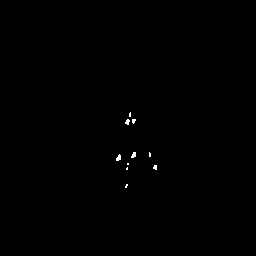} 

  \end{subfigure}
  \hfill 
  \begin{subfigure}{0.325\columnwidth} 
  \includegraphics[width=\textwidth, height=\textwidth]{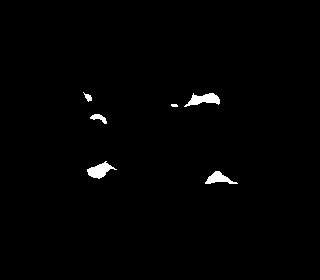} 
 
  \end{subfigure}
  \caption{Example of cross-sectional brain MRI pictures and corresponding masks from an individual with MS }
  \label{fig:test}
 \end{figure}
We have used $2$ publicly available datasets for the experiments. The first dataset \cite{muslim2022brain} comprises MRI scans from $60$ individuals diagnosed with Multiple Sclerosis (MS). Three experts in neurology-radiology conducted the lesions' segmentation manually, ensuring consensus across T1-weighted, T2-weighted and FLAIR MRI sequences. As mentioned, the dataset consists of 3D FLAIR MRI scans obtained from the patients. Among them, $46$ are female and $14$ are male, with an average age of $33$ years, ranging from $15$ to $56$ years. These scans were conducted between $2019$ and $2020$ using a $1.5$ Tesla scanner across $20$ different medical centers. The second dataset \cite{commowick:hal-03358968} consists of FLAIR images from $100$ individuals with MS. These images were obtained from $15$ different MRI scanners, with six operating at $1.5T$ and nine at $3T$. Among these scanners were three from GE, six from Philips and six from Siemens. The images exhibit size and voxel spacing variations, standardized to the dataset's median spacing of $0.977 \times 0.977 \times 0.530 \: mm^3$ before model training. Each patient underwent two scans, spaced $1$–$3$ years apart, resulting in $200$ images. Annotations focused solely on new lesions identified at the second time point, omitting delineation of pre-existing lesions or changes in size. Four neuroradiologists independently annotated each patient, with a consensus that a new lesion mask was generated for model training and evaluation purposes.

\subsection{Data Processing Steps}
The initial processing involved converting volumetric MRI data into individual 2D slices, facilitating focused analysis per slice and isolating potential lesions.

Standardizing the size of 2D slices to $256 \times 256$ pixels was crucial to address variations in image sizes resulting from different acquisition parameters and equipment.

Image sharpening and intensity normalization were integrated into the preprocessing pipeline to enhance MRI data. This technique accentuated edges and subtle abnormalities while maintaining data integrity and mitigating artifacts and noise.

The dataset underwent partitioning, distributing $70\%$ of the images to the training set, while $15\%$ each was designated for validation and testing purposes.

Collaboration with medical professionals specializing in MRI data provided valuable insights into radiological nuances. Their expertise guided preprocessing steps such as noise reduction and intensity normalization, ensuring data reliability for deep learning-based lesion segmentation. This interdisciplinary approach enhances the model's understanding of MRI intricacies and improves segmentation accuracy.

\subsection{Evaluation Metrics}
Within the scope of this study, the evaluation of segmentation algorithms extended to encompass several critical metrics, Intersection over Union ($IoU$), Dice Similarity Coefficient ($DSC$), Hausdorff distance at $95\%$ percentile ($HD_{95}$), $Sensitivity$ and $Specificity$. These metrics collectively provided a comprehensive assessment of the algorithm's performance.

In accordance with assessing whether the model was prone to overfitting, underfitting, or achieving an optimal fit, the results featured the presentation of the $IoU$ for prostate segmentation tasks.
\begin{align*}
    IoU &= \frac{Intersection \: Area}{Union \: Area}\\
        &= \frac{TP}{TP+FP+FN}
\end{align*}
where, Intersection Area is equivalent to True Positive ($TP$) and Union Area is equivalent to $TP$ + False Positive ($FP$) + False Negative ($FN$). 

$DSC$, a metric based on area overlap, quantified the agreement between manual and algorithmic segmentations:

$$DSC = \frac{2TP}{2TP + FP + FN}$$

$HD_{95}$ is a metric used to quantify the similarity between two sets of points, often employed in image processing and computer vision tasks such as image registration or object recognition. It measures the maximum distance from each point in one set to the nearest point in the other set, considering only the closest $95\%$ of points. Mathematically, it can be represented as:
$$HD_{95}(A,B) = max_{a \in A} \{min_{b \in B}\{d(a, b)\}\}$$
where $A$ and $B$ are sets of points, $a$ and $b$ are individual points in $A$ and $B$ respectively and $d(a, b)$ is the distance between point $a$ and point $b$. The $95\%$ indicates that only the distances corresponding to the $95$ percentile of the distances are considered in this calculation.

$Sensitivity$ measures the proportion of actual positive cases that a diagnostic test or model correctly identifies. It is calculated as the ratio of true positive cases to the sum of true positive and false negative cases. Mathematically, $Sensitivity$ can be represented as:
$$Sensitivity = \frac{TP}{TP+FN}$$

$Specificity$ measures the proportion of actual negative cases correctly identified by a diagnostic test or model. It is calculated as the ratio of true negative cases to the sum of true negative and false positive cases. Mathematically, $Specificity$ can be represented as:
$$Specificity  = \frac{TN}{TN+FP}$$
where $TN$ is the number of true negatives.

\section{Result and Discussion}
The evaluation conducted on the test dataset reveals that Mamba-HUNet outperforms all other models across multiple performance metrics. Notably, it achieves an $IoU$ of $0.8536$, a $HD_{95}$ of $2.2518$, a $DSC$ of $0.9287$, a $Sensitivity$ of $0.9294$ and a $Specificity$ of $0.9865$. These results highlight the proposed architecture's superior segmentation accuracy and robustness, indicating its potential for advancing medical image analysis.

The success of Mamba-HUNet can be attributed to its unique design, which integrates features from both Mamba-UNet and HUNet models. By leveraging the linear scaling advantage of Mamba and the global context understanding of HUNet, the architecture effectively captures both local features and long-range dependencies within medical images. Furthermore, the meticulous architecture design, including the Contraction and Expansive Paths, allows for efficient feature extraction and spatial resolution restoration, enhancing segmentation outcomes.

Accurate segmentation of anatomical structures and lesions is crucial for diagnosis and treatment planning in medical imaging. The robust performance of Mamba-HUNet across diverse datasets and complex anatomical structures underscores its potential for real-world applications in clinical settings. Moreover, qualitative results of lesion segmentation demonstrate the model's ability to accurately delineate lesions in medical images, providing further evidence of its efficacy in capturing intricate features and facilitating precise segmentation. Overall, the study's results showcase the effectiveness of Mamba-HUNet as a robust solution for medical image segmentation tasks, with implications for advancing the field of medical image analysis and improving clinical decision-making processes.

\begin{figure}[htb!]
\centering
\begin{tabular}{ccc}
    Test Image & Test Label & Predicted Label \\
    \includegraphics[width=0.30\linewidth]{2022_dataset_1_img.nii_slice_2.png} & \includegraphics[width=0.30\linewidth]{2022_dataset_1_gt.nii_slice_2.png} & \includegraphics[width=0.30\linewidth]{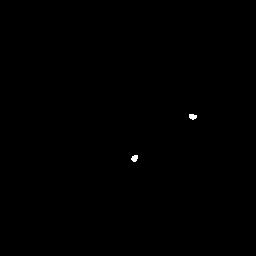} \\
    \includegraphics[width=0.30\linewidth]{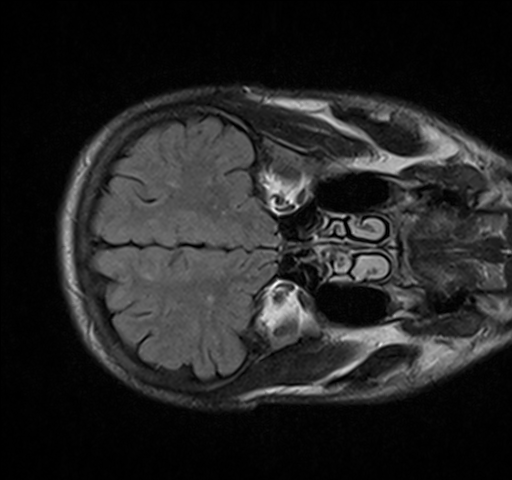} & \includegraphics[width=0.30\linewidth]{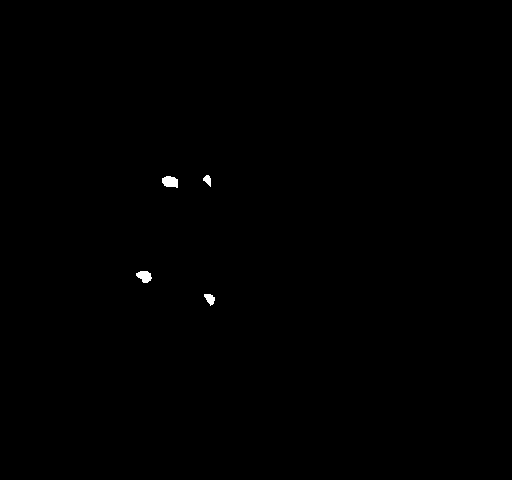} & \includegraphics[width=0.30\linewidth]{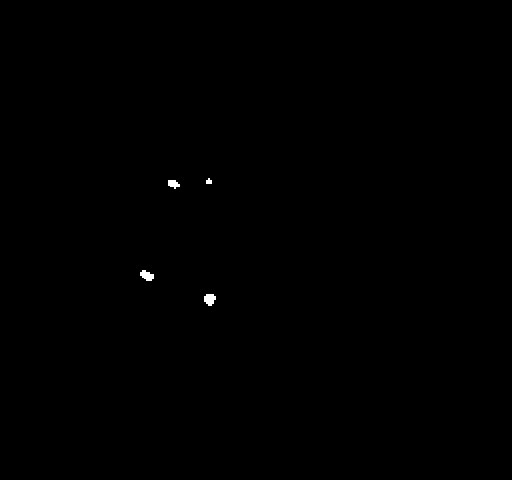} \\
    \includegraphics[width=0.30\linewidth]{2022_dataset_28_img.nii_slice_4.png} & \includegraphics[width=0.30\linewidth]{2022_dataset_28_gt.nii_slice_4.png} & \includegraphics[width=0.30\linewidth]{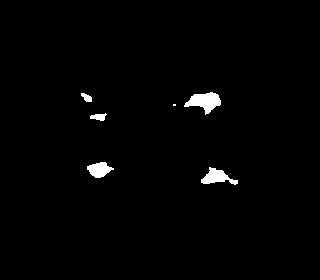} \\
    \includegraphics[width=0.30\linewidth]{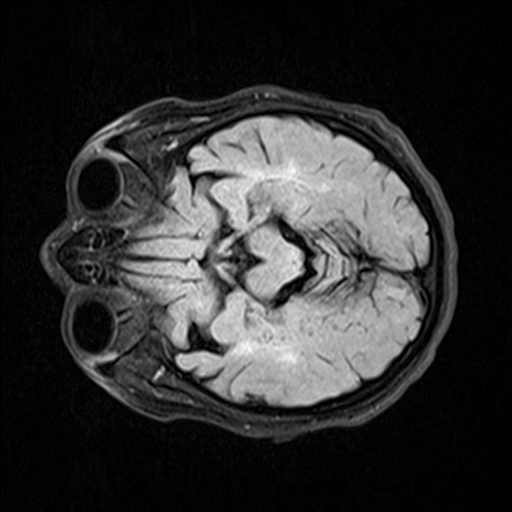} & \includegraphics[width=0.30\linewidth]{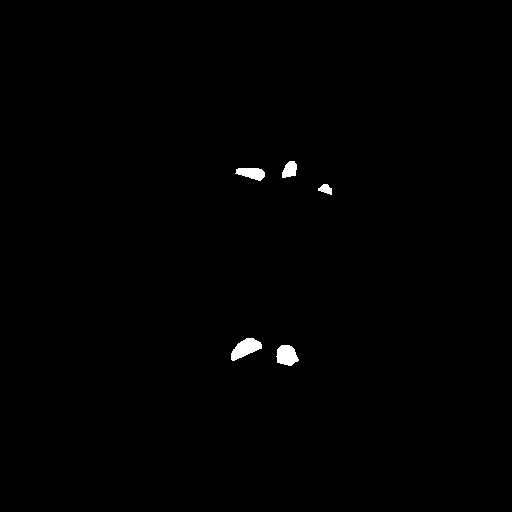} & \includegraphics[width=0.30\linewidth]{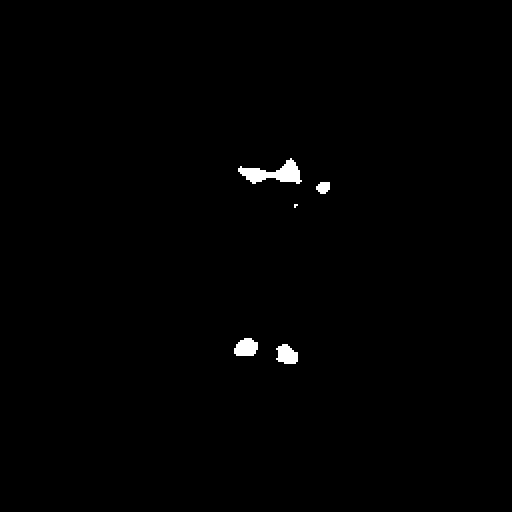}
\end{tabular}
\caption{Output segmentation results of the proposed method}
\label{fig:segmented_results}
\end{figure}

Additionally, Table \ref{fig:lesion_results_via_epoch} summarizes the quantitative performance metrics of various segmentation models, clearly comparing Mamba-HUNet with other state-of-the-art architectures. Furthermore, Figure \ref{fig:segmented_results} showcases sample predicted images with segmentation masks, offering visual insights into the model's segmentation accuracy and demonstrating its potential for clinical applications. These supplementary materials augment the discussion, providing comprehensive evidence of the efficacy and potential impact of Mamba-HUNet in medical image analysis.

\begin{table}
  \centering
  \begin{tblr}{
      colspec={lccccc},
      row{1}={font=\bfseries},
      row{even}={bg=gainsboro229},
    }
    \toprule
    Model & IoU$\uparrow$ & HD$_{95}$ $\downarrow$ & DSC$\uparrow$ & Sensitivity$\uparrow$ & Specificity$\uparrow$\\
    \toprule
    U-Net & $0.8170$ & $3.2158$ & $0.8597$ & $0.9045$ &  $0.9395$\\
    Swin-Unet & $0.7947$ & $4.5927$ & $0.8437$ & $0.8921$ &  $0.9186$\\
    Mamba-UNet & $0.8322$ & $2.3875$ & $0.8761$ & $0.9178$ &  $0.9787$\\
    VM-Unet & $0.8039$ & $2.9006$ & $0.8749$ & $0.9135$ &  $0.9531$\\
    \textbf{Mamba-HUNet} & \textbf{$0.8536$} & \textbf{$2.2518$} & \textbf{$0.9287$} & \textbf{$0.9294$} &  \textbf{$0.9865$}\\
    \bottomrule
  \end{tblr}
  \caption{Performance of different models on the test dataset. $\uparrow$ denotes higher is better and $\downarrow$ denotes lower is better}
  \label{fig:lesion_results_via_epoch}
\end{table}

\section{Conclusion}
This paper introduced Mamba-HUNet, a novel architecture for robust and efficient medical image segmentation tasks. By integrating components from CNNs and SSMs, Mamba-HUNet harnesses the strengths of both approaches to achieve superior segmentation accuracy and robustness.

Inspired by Mamba-UNet and HUNet models, the architecture combines CNNs' local feature extraction power with SSMs' long-range dependency modeling capabilities. Through careful design, including partitioning input images into patches and leveraging VSS blocks, Mamba-HUNet effectively captures both local details and global contexts within medical images. Skip connections and HUNet-inspired upsampling blocks enhance feature blending and spatial resolution restoration, contributing to superior segmentation outcomes.

Evaluation of MRI scans, particularly in Multiple Sclerosis lesion segmentation, demonstrates Mamba-HUNet's effectiveness across diverse datasets and complex anatomical structures. The model outperforms state-of-the-art architectures, achieving remarkable performance metrics such as $IoU$, $DSC$, $HD_{95}$, $Sensitivity$ and $Specificity$. These results underscore the potential of Mamba-HUNet in advancing medical image analysis, with implications for improving clinical decision-making processes.

The success of Mamba-HUNet lies in its superior performance and potential for real-world applications in clinical settings. Accurate segmentation of anatomical structures and lesions is crucial for diagnosis and treatment planning and the architecture demonstrates promising results in this regard. Furthermore, qualitative assessments highlight the model's ability to accurately delineate lesions in medical images, further affirming its efficacy in capturing intricate features and facilitating precise segmentation.



\bibliography{bibliography}
\bibliographystyle{IEEEtran}
\end{document}